\documentclass[a4paper,12pt]{article}

\usepackage[utf8]{inputenc}  
\usepackage[T1]{fontenc}
\usepackage{amsmath,amssymb}
\usepackage{authblk}

\usepackage[margin=2.5cm]{geometry}
\usepackage{lmodern}
\usepackage{siunitx}
\usepackage{graphicx}
\usepackage{caption}
\usepackage{comment}
\usepackage{float}
\usepackage{tabularx}

\usepackage{xcolor}
\usepackage{ulem}

\usepackage{booktabs}

\usepackage[numbers,sort&compress]{natbib}

\usepackage{tcolorbox}
\tcbuselibrary{listings,breakable}

\newtcolorbox{conditionsbox}[1][]{%
  breakable,
  colback=gray!5,
  colframe=black,
  fonttitle=\bfseries,
  title=#1
}

\newcounter{derivationstep}
\newenvironment{derivationstep}[1][]{\refstepcounter{derivationstep}\par\noindent
   \textbf{\textcolor{blue}{\uline{Step~\thederivationstep: #1}} }}{\par}

\newcounter{condition}

\usepackage{fancyhdr}

\usepackage[hidelinks]{hyperref}

\title{\textbf{From Étendue to the Lowest Fundamental SNR: Pixel Étendue (Optogeometric Factor) Interpreted as Mode Count}}

\author[1]{Jan Sova}
\author[1]{Marie Kolaříková}
\affil[1]{Czech Technical University in Prague, Faculty of Mechanical Engineering}
\date{\today}

\begin{document}
\maketitle

\begin{abstract}
The optogeometric factor $F_{\mathrm{opg}}$, recently introduced as a pixel-level 
form of étendue, quantifies the spatial--angular throughput of a detector element. 
In this work its interpretation is extended by identifying $F_{\mathrm{opg}}$ with the 
number of accessible optical modes per pixel. This mode-based perspective establishes 
a direct link between radiometric throughput and quantum photon statistics. 
By combining $F_{\mathrm{opg}}$ with the Bose--Einstein distribution, an estimate of the 
lowest achievable signal-to-noise ratio (SNR) at the pixel level is derived. 
Explicit formulas are presented in both scene-based and sensor-based forms, showing how 
the minimal SNR depends on aperture geometry, pixel pitch, f-number, wavelength, 
and source temperature. This formulation provides a compact and physically transparent 
benchmark for evaluating imaging sensors against the lowest expected quantum noise limit.
\end{abstract}

\section{Introduction}
In classical infrared imaging, the radiant flux $\Phi_{\mathrm{pix}}\,[\si{W}]$ incident
on a single detector pixel can be expressed as~\cite{opg2025}

\begin{equation}
\Phi_{\mathrm{pix}} = L \, F_{\mathrm{opg}},
\end{equation}

where $L$ denotes the average radiance of the scene 
$[\si{W\,m^{-2}\,sr^{-1}}]$, assumed approximately constant 
over the projected footprint and the pixel solid angle so that it can be factored out 
of the radiometric integral~\cite{hs2025,boyd1983radiometry,grant2011field}. 
The factor $F_{\mathrm{opg}} \,[\si{m^2 sr}]$ is the optogeometric factor,
\textbf{representing the spatial--angular throughput of the optical system per pixel, 
equivalent to the one pixel étendue}.

While this relation is standard in radiometry, it can be given a deeper
interpretation if one considers the quantum structure of the electromagnetic field.
Each mode of the field corresponds to a quantum harmonic oscillatoroscillator~\cite{mandel1995optical,scully1997quantum} 
and, as will be shown below, $F_{\mathrm{opg}}$ \textbf{determines how many such modes are coupled to a single pixel of a camera detector}.

\section{Optogeometric factor as a mode counter}
In wave optics, the number of independent spatial--angular modes supported 
by an optical system is classically expressed as the ratio of its étendue to the 
fundamental phase--space cell of size $\lambda^2$~\cite{mandel1995optical,scully1997quantum}:
\begin{equation}
N_{\mathrm{modes}} \;\approx\; \frac{G}{\lambda^2},
\end{equation}
where $G\,[\si{m^2\,sr}]$ is the system étendue and 
$\lambda\,[\si{m}]$ the optical wavelength, which sets the diffraction--limited 
phase--space resolution per polarization state.

At the level of a single detector element, however, the relevant throughput 
is not the global étendue $G$ but the \emph{optogeometric factor} $F_{\mathrm{opg}}$, 
which quantifies the area--solid--angle acceptance of one pixel. 
Introducing an \emph{effective coherence scale} $\lambda_{\mathrm{pix}}$, 
we generalize the same mode--étendue relation down to the pixel scale:
\[
N_{\mathrm{modes,pix}} \;\approx\; \frac{F_{\mathrm{opg}}}{\lambda_{\mathrm{pix}}^2}.
\]
This refinement preserves the well--known connection between étendue and mode 
count, while making explicit how many quantum degrees of freedom each pixel can 
access.

1. \textbf{Diffraction--limited regime}:  
   If the pixel pitch is much smaller than the diffraction spot size, 
   $a_{\mathrm{pix}} \ll 1.22\,\lambda\,f/\#$, then the resolution is 
   set by the diffraction cell of the optics and
   \[
   \lambda_{\mathrm{pix}} \;\approx\; 1.22\,\lambda\,f/\#.
   \]

where
\begin{itemize}
  \item $\lambda\;[\si{m}]$ is the acoustic (or optical) wavelength in the propagation medium,
  \item $f/\#\;[\si{-}]$ is the f-number of the focusing system, defined as $f/D$ with $f$ the focal length and $D$ the aperture diameter,
  \item $\lambda_{\mathrm{pix}}\;[\si{m}]$ is the effective coherence scale (diffraction-limited spot size) in the detector plane.
\end{itemize} 

2. \textbf{Geometry--limited regime}:  
   If the pixel pitch is large compared to the diffraction spot, 
   $a_{\mathrm{pix}} \gg 1.22\,\lambda\,f/\#$, then the pixel itself limits 
   the resolution and
   \[
   \lambda_{\mathrm{pix}} \;\approx\; a_{\mathrm{pix}}.
   \]

where $a_{\mathrm{pix}}\,[\si{m}]$ denotes the pixel pitch (center–to–center spacing of pixels in the detector array).

These two cases can be compactly combined into the definition
\begin{equation}
\lambda_{\mathrm{pix}} := \max\!\left( 1.22\,\lambda\,f/\#, \; a_{\mathrm{pix}} \right),
\end{equation}
which expresses that the effective coherence scale is always determined by 
the larger of the diffraction blur and the projected pixel pitch. 
This generalization makes explicit the dual role of $F_{\mathrm{opg}}$: 
it bridges system étendue with pixel--level mode theory, clarifying how many 
quantum degrees of freedom each detector element can access.

This generalization makes explicit the dual role of $F_{\mathrm{opg}}$: 
it bridges system étendue with pixel--level mode theory, 
clarifying how many quantum degrees of freedom each detector element can access. 

From this perspective, each pixel can be regarded as admitting only a 
finite number of independent spatial--angular modes. 
The optogeometric factor $F_{\mathrm{opg}}$ thus plays the role of a 
``mode allocator'' at the pixel level: dividing $F_{\mathrm{opg}}$ by the 
effective phase--space cell $\lambda_{\mathrm{pix}}^2$ yields the exact mode 
count per pixel. This interpretation establishes the ground for the 
oscillator formalism developed in the next section.

The optogeometric factor, which represents the étendue per pixel, 
is rigorously defined~\cite{opg2025} by the surface–solid–angle integral

\begin{equation}
F_{\mathrm{opg}}
\;:=\;
\iint_{A_{\mathrm{fp}}}
\iint_{\Omega_{\mathrm{pix}}(\mathbf{r})}
\cos\theta\,\mathrm d\Omega\,\mathrm dA,
\label{eq:Fopg_def}
\end{equation}

where $A_{\mathrm{fp}}$ is the (unprojected) footprint in the scene,
$\Omega_{\mathrm{pix}}(\mathbf{r})$ is the solid angle of acceptance of the pixels as seen from the point $\mathbf{r}$
on the footprint, and $\theta$ is the local incidence angle between the normal surface at $\mathbf{r}$
and the viewing direction. 
This definition is meaningful only under the assumption of uniform radiance in the pixel footprint and angular cone, so that the radiance of the scene $L$ can be taken out of the double integral. 
The quantity $F_{\mathrm{opg}}$ then has units $[\si{m^2\,sr}]$~\cite{opg2025}, that is, it represents an area–solid–angle volume element in the optical phase space~\cite{boyd1983radiometry,grant2011field,ISO9334}.

Under paraxial conditions with almost normal incidence (so that $\cos\theta\!\approx\!1$) 
and weak spatial variation of $\Omega_{\mathrm{pix}}(\mathbf{r})$, Eq.~\eqref{eq:Fopg_def} reduces to
\begin{equation}
F_{\mathrm{opg}} \;\approx\; A_{\mathrm{fp}}^{*}\,\Omega_{\mathrm{pix}},
\label{eq:Fopg_parax}
\end{equation}
where $A_{\mathrm{fp}}^{*}$ is the projected footprint area $[\si{m^2}]$, i.e.\ the 
projection of the scene footprint onto the plane of the detector pixel (assumed planar), 
and $\Omega_{\mathrm{pix}}$ the (approximately constant) pixel solid angle $[\si{sr}]$.

In a recent study~\cite{hs2025}, the optogeometric factor appeared in a reduced and 
approximate form, denoted as $A_{\mathrm{x}}\;[\si{\meter^{2}}]$, which was introduced 
to simplify comparison with the conceptual thermography equation. 
In the present work, we adopt the more general notation 
$\tilde{\bar{F}}_{\mathrm{opg}}^{(D,\varphi)}$, with the following relation:

\begin{equation}
\pi A_{\mathrm{x}}\;[\si{\meter^{2}\,sr}]
\;\equiv\; 
\pi\,\tilde{\bar{F}}_{\mathrm{opg}}^{(D,\varphi)} ,
\label{eq:Ax_relation}
\end{equation}

showing the correspondence between the previously used reduced notation and the present formulation.

As derived in~\cite{opg2025}, the optogeometric factor can be expressed either on the sensor side or on the scene side, as follows:

\begin{align}
\tilde{\bar{F}}_{\mathrm{opg}}^{(D,\varphi)} & = \tfrac{1}{4}\,D^{2}\,\varphi_{\mathrm{iFOV}}^{2}, 
& \text{(reduced scene--based)}, \label{eq:Fopg_scene} \\[0.3em]
\tilde{\bar{F}}_{\mathrm{opg,s}}^{(a,f\#)} & = \tfrac{1}{4}\,\Bigl(\tfrac{a}{f\#}\Bigr)^{2}, 
& \text{(reduced sensor--based)}, \label{eq:Fopg_sensor}
\end{align}

where $D$ is the entrance pupil diameter $[\si{m}]$, 
$\varphi_{\mathrm{iFOV}}$ the pixel instantaneous field of view $[\si{rad}]$, 
$a$ the pixel pitch $[\si{m}]$, 
and $f\#=f/D$ the f--number of the lens $[\si{-}]$. The thermography equation formulated using the scene-based and sensor-based optogeometric factor, as a practical demonstration of its application, is presented in~\cite{sova2025thermography}.

\section{Quantum oscillator definition}
Each optical mode of the electromagnetic field is mathematically equivalent 
to a quantum harmonic oscillator with discrete energy levels

\begin{equation}
E_n = \hbar \,\frac{2\pi c}{\lambda_{\mathrm{osc}}}\,\left(n + \tfrac{1}{2}\right),
\end{equation}

where
\begin{itemize}
    \item $E_n \;[\si{\joule}]$ is the energy of the $n$--th level,
    \item $\hbar = 1.054\times 10^{-34}\;[\si{\joule\second}]$ is the reduced Planck constant,
    \item $c = 2.998\times 10^{8}\;[\si{\meter\per\second}]$ is the speed of light in vacuum,
    \item $\lambda_{\mathrm{osc}}\;[\si{\meter}]$ is the wavelength corresponding to the oscillator frequency,
    \item $n \in \{0,1,2,\dots\}$ is the mode quantum number (dimensionless).
\end{itemize}

The fundamental unit in this relation is the minimum phase–space area 
of a single optical mode, given by $\lambda^{2}\,[\si{m^2}]$, 
corresponding to one spatial–angular degree of freedom per polarization state~\cite{mandel1995optical,scully1997quantum}. 
When explicitly normalized per pixel, we denote this unit as $\lambda_{\mathrm{pix}}^2$.

Since the optogeometric factor $F_{\mathrm{opg}}$ represents the pixel étendue, 
the number of independent optical oscillators (modes) geometrically admitted by one pixel is
\begin{equation}
N_{\mathrm{osc}} = \frac{F_{\mathrm{opg}}}{\lambda_{\mathrm{pix}}^2}.
\label{eq:Nosc}
\end{equation}

If $N_{\mathrm{osc}} \geq 1$, the pixel can couple to one or more independent spatial–angular modes, 
each corresponding to a quantum harmonic oscillator.  
If $N_{\mathrm{osc}} < 1$, the throughput of the pixel is smaller than a single phase–space cell, 
so the detector effectively admits only a fractional occupancy of one mode.

This relation provides a direct bridge between classical radiometry and 
quantum optics: $F_{\mathrm{opg}}$ measures the geometric throughput, while 
$N_{\mathrm{osc}}$ expresses the same quantity in quantum terms as a mode count. 
If $N_{\mathrm{osc}}=2$, the pixel accepts two independent oscillators; 
if $N_{\mathrm{osc}}<1$, the pixel throughput corresponds to a fraction 
of a full coherence cell.

Using the paraxial form of the optogeometric factor, a representative 
LWIR pixel with $a_{\mathrm{pix}}=\SI{17}{\micro\meter}$ and $f/\#=1.0$ 
gives $\tilde F_{\mathrm{opg}} \approx \SI{2.27e-10}{m^2\,sr}$. 
At wavelength $\lambda=\SI{10}{\micro\meter}$,
\[
N_{\mathrm{osc}} \;=\; \frac{2.27\times 10^{-10}}{(10^{-5})^2} \;\approx\; 2.27,
\]
that is, only a few independent modes per pixel for the representative LWIR case.

This identification of $F_{\mathrm{opg}}$ with a normalized mode count provides the 
missing link: a purely geometric radiometric quantity is reinterpreted as a 
quantum degree-of-freedom count, thereby establishing a rigorous bridge between 
classical radiometry and quantum optics. 

The novelty of the present work lies in translating this well-known 
mode–étendue correspondence, usually formulated at the system level, 
down to the scale of a single detector pixel. By explicitly linking the 
radiometric throughput $F_{\mathrm{opg}}$ to the quantum oscillator 
normalization, we obtain compact pixel-level expressions for photon statistics 
and the fundamental signal-to-noise ratio (SNR). This pixel-based reformulation 
is absent in prior literature and provides a practical benchmark for the design 
and evaluation of real imaging sensors, where the limiting sensitivity is set by 
the finite number of optical modes that each pixel can admit.

\begin{table}[H]
\centering
\caption{Oscillator (mode) count per pixel for 
$\tilde F_{\mathrm{opg}}=\SI{2.27e-10}{m^2\,sr}$.}
\label{tab:Nosc_vs_lambda}
\begin{tabular}{rcc}
\hline
$\lambda$ & $\lambda^2$ & $N_{\mathrm{osc}} = F_{\mathrm{opg}}/\lambda^2$ \\
\hline
\SI{1}{\micro\meter}  & $1.0\times 10^{-12}\ \si{m^2}$  & $2.27\times 10^{2}$ \\
\SI{3}{\micro\meter}  & $9.0\times 10^{-12}\ \si{m^2}$  & $2.52\times 10^{1}$ \\
\SI{5}{\micro\meter}  & $2.5\times 10^{-11}\ \si{m^2}$  & $9.08$ \\
\SI{10}{\micro\meter} & $1.0\times 10^{-10}\ \si{m^2}$ & $2.27$ \\
\SI{14}{\micro\meter} & $1.96\times 10^{-10}\ \si{m^2}$ & $1.16$ \\
\hline
\end{tabular}
\end{table}

This compact definition clarifies that a pixel is not merely a geometric 
collecting area, but an aperture to a finite number of quantum oscillators. 
It thus sets the stage for analyzing photon statistics and the 
quantum--limited SNR in the following section.

\section{Implications for photon statistics and SNR}
The average photon occupancy of a single optical mode at frequency 
$\lambda_{\mathrm{meas}}$ and temperature $T$ follows from Bose--Einstein statistics 
\cite{mandel1995optical,scully1997quantum}:

\begin{equation}
\bar n(\lambda_{\mathrm{meas}},T)
= \frac{1}{\exp\!\left(\tfrac{hc}{\lambda_{\mathrm{meas}} kT}\right) - 1},
\end{equation}

where $\lambda_{\mathrm{meas}}$ denotes the measurement wavelength corresponding 
to the photon frequency $\nu = c/\lambda_{\mathrm{meas}}$. 
Here $h$ is Planck’s constant ($\SI{6.62607015e-34}{J\,s}$), 
$k$ is Boltzmann’s constant ($\SI{1.380649e-23}{J\,K^{-1}}$), 
and $c$ is the speed of light in vacuum ($\SI{2.99792458e8}{m\,s^{-1}}$). 

In general, the structure of the quantum-limited SNR remains unchanged; 
only the modal occupancy $\bar n$ must be replaced by the appropriate 
statistical factor depending on the photon source 
(Bose–Einstein for thermal radiation, Poisson for coherent states, 
or fixed $n$ for Fock states).

\begin{table}[H]
\centering
\caption{Two distinct uses of wavelength: measurement wavelength for photon statistics vs.\ geometric coherence scale for mode counting.}
\label{tab:lambdas}
\begin{tabularx}{\linewidth}{>{\hsize=0.16\hsize}l
                                  >{\hsize=0.34\hsize}X
                                  >{\hsize=0.50\hsize}X}
\toprule
\textbf{Symbol} & \textbf{Definition} & \textbf{Role in this work} \\
\midrule
$\lambda_{\mathrm{meas}}$ 
& \textbf{Measurement wavelength} (photon wavelength within the detection band), with $\nu = c/\lambda_{\mathrm{meas}}$ 
& Enters the Bose--Einstein occupancy 
$\displaystyle \bar n(\lambda_{\mathrm{meas}},T)=\bigl[\exp\!\bigl(\tfrac{hc}{\lambda_{\mathrm{meas}}kT}\bigr)-1\bigr]^{-1}$, 
thereby linking the source temperature $T$ to the mean photon number per spatial–angular mode. 
Chosen by the detector/filter band. \\[0.6em]
$\lambda_{\mathrm{pix}}$ 
& \textbf{Geometric coherence scale} in the detector plane,
$\displaystyle \lambda_{\mathrm{pix}} := \max\!\bigl(1.22\,\lambda_{\mathrm{meas}}\,f/\#,\, a_{\mathrm{pix}}\bigr)$
& Sets the phase–space cell size for mode counting, 
$\displaystyle N_{\mathrm{osc}} = F_{\mathrm{opg}}/\lambda_{\mathrm{pix}}^{2}$. 
Determined by optics ($f/\#$) and pixel pitch $a_{\mathrm{pix}}$; independent of the source temperature. \\
\bottomrule
\end{tabularx}
\end{table}

The total photon number collected by a pixel then follows
from those steps:

\begin{derivationstep}[Average modal occupancy (Bose--Einstein)]\label{step:modal_occupancy}
Each spatial--angular mode of the electromagnetic field at thermal equilibrium 
carries on average $\bar n(\lambda_{\mathrm{meas}},T)$ photons, given by the Bose--Einstein distribution.
\end{derivationstep}

\begin{derivationstep}[Mode count from optogeometric factor]\label{step:mode_count}
A detector pixel does not couple to a single mode, but to a finite number of
independent modes determined by its geometric throughput. As established above,
the number of such oscillators (modes) admitted by a pixel is
\begin{equation}
N_{\mathrm{osc}} = \frac{F_{\mathrm{opg}}}{\lambda_{\mathrm{pix}}^{2}}.
\end{equation}
\end{derivationstep}

\begin{derivationstep}[Mode count × average occupancy]\label{step:modes_times_occupancy}
We define the \emph{effective} number of spatial–angular modes collected by a pixel 
during the integration time $\tau$ and within the detection bandwidth $\Delta\nu$ as
\begin{equation}
N_{\mathrm{modes}}^{\mathrm{eff}}
= \eta_{\mathrm{sys}}\,N_{\mathrm{pol}}\,
\frac{F_{\mathrm{opg}}}{\lambda_{\mathrm{pix}}^{2}}\,
(\Delta\nu\,\tau),
\end{equation}
where $\eta_{\mathrm{sys}}$ is the system efficiency and $N_{\mathrm{pol}}$ the number of 
polarization states.

The corresponding photon number estimate can then be written in the compact form
\begin{equation}
N_{\mathrm{ph}} \;\approx\; N_{\mathrm{modes}}^{\mathrm{eff}}\;\bar n(\lambda_{\mathrm{meas}},T),
\end{equation}
that is, ``\emph{number of accessible modes} × \emph{average photon occupancy per mode}''. 

Here $\lambda_{\mathrm{meas}}$ denotes the measurement wavelength at which the photon 
energy $E=hc/\lambda_{\mathrm{meas}}$ and the Bose–Einstein occupancy 
$\bar n(\lambda_{\mathrm{meas}},T)$ are evaluated, while $\lambda_{\mathrm{pix}}$ defines the 
geometric coherence scale for mode counting.
\end{derivationstep}

\begin{derivationstep}[Quantum noise and shot-noise limit]\label{step:snr_shot}
Photon detection is subject to quantum fluctuations arising from the discrete 
nature of photons. In the shot-noise limit, the variance of the detected photon 
number equals the mean value,
\begin{equation}
\sigma_N^2 \;=\; N_{\mathrm{ph}},
\end{equation}
so that the noise amplitude is
\begin{equation}
\sigma_N = \sqrt{N_{\mathrm{ph}}}.
\end{equation}
Accordingly, the fundamental signal-to-noise ratio is

\begin{equation}
\mathrm{SNR}_{\mathrm{fund}} \;:=\; \frac{N_{\mathrm{ph}}}{\sigma_N} \;=\; \sqrt{N_{\mathrm{ph}}}.
\end{equation}

This represents the quantum-limited case; any additional detector or background 
noise sources can only reduce the SNR further.
\end{derivationstep}

\begin{derivationstep}[Conditions for the compact form of SNR]
Starting from
\[
N_{\mathrm{ph}} \;\approx\;
\eta_{\mathrm{sys}}\,N_{\mathrm{pol}}\,
\frac{F_{\mathrm{opg}}}{\lambda_{\mathrm{pix}}^{2}}\,
(\Delta\nu\,\tau)\,
\bar n(\lambda_{\mathrm{meas}},T),
\qquad
\mathrm{SNR}_{\mathrm{fund}}=\sqrt{N_{\mathrm{ph}}},
\]
the simplified form
\begin{equation}
\mathrm{SNR}_{\mathrm{fund}} \;\approx\; 
\sqrt{ \frac{F_{\mathrm{opg}}}{\lambda_{\mathrm{pix}}^{2}}\,\bar n(\lambda_{\mathrm{meas}},T) }
\label{eq:SNF_fund_aprox}
\end{equation}

follows under the assumptions:

\begin{enumerate}\itemsep0.3em
\item \textbf{Narrowband \& uniform scene:} Radiance is approximately constant over the pixel footprint and acceptance cone (paraxial, near-normal incidence), so $F_{\mathrm{opg}}$ is well-defined.  
\item \textbf{Shot-noise limit:} Photon-number fluctuations follow $\sigma_N^2\simeq N_{\mathrm{ph}}$, i.e.\ quantum noise dominates.  
\item \textbf{Negligible technical noise:} Detector/readout noise and background contributions are ignored.  
\item \textbf{Normalized system factors:} $\eta_{\mathrm{sys}}=1$, $N_{\mathrm{pol}}=1$, and $\Delta\nu\,\tau=1$ (absorbed into the occupancy for this compact form).  
\end{enumerate}
\end{derivationstep}

\subsection{SNR expressed via scene-based and sensor-based optogeometric factors}

In order to obtain explicit pixel-level formulas, we substitute the 
scene-based and sensor-based reduced forms of the optogeometric factor, 
Eqs.~\eqref{eq:Fopg_scene}--\eqref{eq:Fopg_sensor}, 
into the general expression for the fundamental signal-to-noise ratio, 
Eq.~\eqref{eq:SNF_fund_aprox}. 
This yields two compact formulations of $\mathrm{SNR}$ directly in terms of 
optical aperture geometry (scene-based) or sensor design parameters (sensor-based).

\paragraph{Scene-based form.}
With $\pi\tilde{\bar{F}}_{\mathrm{opg}}^{(D,\varphi)} = \tfrac{\pi}{4}D^{2}\varphi_{\mathrm{iFOV}}^{2}$, 
Eq.~\eqref{eq:SNF_fund_aprox} gives
\begin{equation}
\mathrm{SNR}_{\mathrm{fund}}
\;\approx\;
\frac{D\,\varphi_{\mathrm{iFOV}}}{2}\;
\sqrt{\frac{\pi}{\lambda_{\mathrm{pix}}^{2}}\,\bar n(\lambda_{\mathrm{meas}},T)}.
\end{equation}

\paragraph{Sensor-based form.}
With $\pi\tilde{\bar{F}}_{\mathrm{opg,s}}^{(a,f\#)} = \tfrac{\pi}{4}\bigl(\tfrac{a}{f\#}\bigr)^{2}$, 
Eq.~\eqref{eq:SNF_fund_aprox} gives
\begin{equation}
\mathrm{SNR}_{\mathrm{fund}}
\;\approx\;
\frac{a_{\mathrm{pix}}}{2\,f\#}\;
\sqrt{\frac{\pi}{\lambda_{\mathrm{pix}}^{2}}\,\bar n(\lambda_{\mathrm{meas}},T)}.
\end{equation}

\noindent
Both expressions make explicit how the pixel-level SNR is governed by either 
aperture geometry or pixel design, while the modal occupancy 
$\bar n(\lambda_{\mathrm{meas}},T)$ remains set by the photon statistics 
of the source.

\begin{tcolorbox}[colback=gray!5,colframe=black,title={Fundamental pixel-level SNR}]
The signal--to--noise ratio of any real imaging system at the level of a single detector pixel
is fundamentally limited by photon statistics. 
For a pixel of area $a_{\mathrm{pix}}^2$ and optics of f-number $f\#$, the quantum--limited 
signal--to--noise ratio is
\begin{equation}
\mathrm{SNR}_{\mathrm{fund}}
= \sqrt{ \frac{F_{\mathrm{opg}}}{\lambda_{\mathrm{pix}}^{2}}\,\bar n( \lambda_{\mathrm{meas}},T) }
\;\approx\;
\frac{a_{\mathrm{pix}}}{2\,f\#}\;
\sqrt{\frac{\pi}{\lambda_{\mathrm{pix}}^{2}}\,\bar n( \lambda_{\mathrm{meas}},T)}.
\label{eq:SNR_final}
\end{equation}
In practice,
\begin{equation}
\mathrm{SNR}_{\mathrm{real}} \;<\; \mathrm{SNR}_{\mathrm{fund}},
\end{equation}
since additional detector noise sources (Johnson noise, $1/f$ noise, dark current, readout 
electronics) can only degrade the performance further. 
Equation above therefore represents the ultimate benchmark for SNR at the pixel level.
\end{tcolorbox}

It follows directly from Eq.~\eqref{eq:SNR_final} that the fundamental 
signal-to-noise ratio depends not only on the optical throughput and pixel 
geometry, but also on the source temperature $T$ and the measurement 
wavelength $\lambda_{\mathrm{meas}}$ through the modal occupancy 
$\bar n(\lambda_{\mathrm{meas}},T)$. 
At longer wavelengths the photon energy decreases, which increases 
$\bar n$ and therefore improves the achievable SNR, whereas at shorter 
wavelengths the modal occupancy becomes small and the SNR is correspondingly 
reduced. 
Thus, the quantum-limited performance of an imaging system is inherently 
tied to both the spectral band of operation and the physical temperature of 
the observed scene.

\section{Conclusion}
The optogeometric factor $F_{\mathrm{opg}}$ has been reinterpreted as a mode counter,
establishing a direct connection between pixel étendue and the number of
quantum harmonic oscillators admitted by a detector pixel. This represents
the first explicit formulation that makes the quantum nature of pixel étendue
mathematically transparent, linking geometric throughput to quantum degrees
of freedom.

From this perspective, $F_{\mathrm{opg}}$ sets the ultimate benchmark for
photon statistics and quantum--limited SNR at the pixel level. The result
is a compact and physically grounded framework that unifies classical
radiometry with quantum optics. This conceptual shift clarifies the role of
geometry in photon detection and provides practical criteria for evaluating
and designing next--generation imaging systems.

For thermal radiation, the Bose--Einstein distribution represents the maximum-entropy case, 
which sets the lowest possible SNR at a given photon flux. 
Non-thermal sources (e.g.\ lasers or engineered photon states) exhibit reduced entropy 
and correspondingly higher SNR, since their photon statistics deviate from the Bose--Einstein limit.

\bibliographystyle{plain}

\begin{thebibliography}{99}
\bibitem{hs2025}
M. Hofreiter and J. Sova and M. Kolaříková and L. Kolařík and T. Němec, \textit{Ther\-mo\-gra\-phy Equation for Non-Per\-pen\-dic\-u\-lar Infrared Mea\-sure\-ments: Der\-i\-va\-tion, Anal\-y\-sis, and Ex\-per\-i\-men\-tal Val\-i\-da\-tion}, Infrared Physics and Technology, 2025, p.~106010, doi:\href{https://doi.org/10.1016/j.infrared.2025.106010}{10.1016/j.infrared.2025.106010}.

\bibitem{opg2025}
Sova, Jan and Kolaříková, Marie, 
\textit{Geometric Optical Throughput and Etendue Revisited: Introducing the Optogeometric Factor for Pixel Level Quantitative Imaging}, 
arXiv preprint arXiv:2508.09335, 2025, 
doi:\href{https://doi.org/10.48550/arXiv.2508.09335}{10.48550/arXiv.2508.09335}.

\bibitem{sova2025thermography}
Sova, Jan and Kolaříková, Marie,
\textit{Thermography Equation: From Conceptual Relation to Quantitative Formulation via the Optogeometric Factor},
arXiv preprint arXiv:2508.11455, 2025,
doi:\href{https://doi.org/10.48550/arXiv.2508.11455}{10.48550/arXiv.2508.11455}.


\bibitem{mandel1995optical}
Mandel, Leonard and Wolf, Emil, 
\textit{Optical Coherence and Quantum Optics}, 
Cambridge University Press, Cambridge, 1995, 
doi:\href{https://doi.org/10.1017/CBO9781139644105}{10.1017/CBO9781139644105}.

\bibitem{scully1997quantum}
Scully, Marlan O. and Zubairy, M. Suhail, 
\textit{Quantum Optics}, 
Cambridge University Press, Cambridge, 1997, 
doi:\href{https://doi.org/10.1017/CBO9780511813993}{10.1017/CBO9780511813993}.


\bibitem{ISO9334}
International Organization for Standardization, \textit{ISO 9334 - Optics and photonics — Optical transfer function — Definitions and mathematical relationships}, 2012,

\bibitem{grant2011field}
Grant, Barbara G., \textit{Field Guide to Radiometry}, 2011.

\bibitem{boyd1983radiometry}
Robert W. Boyd, \textit{Radiometry and the Detection of Optical Radiation}, 1983.

\end{thebibliography}

\end{document}